\documentclass[12pt]{iopart}

\usepackage{graphicx}
\usepackage{textcomp}
\usepackage{float}

\usepackage{amssymb}
\usepackage{subfigure}

\begin{document}

\title[The effect of electron scattering on thermoelectric transport in CoSi]{The effect of energy-dependent electron scattering on thermoelectric transport in novel topological semimetal CoSi.}
\author{D A Pshenay-Severin$^1$, Y V Ivanov$^1$, A T  Burkov$^1$}
\address{$^1$Ioffe Institute, Saint Petersburg 194021, Russia}
\date{\today}
\begin{abstract}Cobalt monosilicide and its solid solutions with Fe or Ni crystallize in B20 cubic noncentrosymmetric structure.
These compounds have long been known as promising thermoelectric materials. 
Recently it was revealed, that they also have unconventional electronic topology.
This renewed interest to the investigation of their transport properties. 
In order to improve theoretical description of thermoelectric transport in these compounds, we take into account electron scattering beyond commonly used constant relaxation time approximation. 
Using first principle calculations, we investigate the scattering of charge carriers by phonons and point defects.
The dependence of the scattering rate on the energy correlates with that for the total density of states. 
This implies that in this material not only the intraband, but also the interband scattering is important, especially for bands with low density of states. 
The Seebeck coefficient and the electrical resistivity of CoSi and of dilute solid solutions Co$_{1-x}$M$_x$Si (M=Fe or Ni, $x<0.1$) are calculated as a function of temperature and the alloy composition. 
We show that the account of strong energy dependence of relaxation time is important for the description of experimentally observed rapid increase of the resistivity and qualitative change of its temperature dependence with the substitution of cobalt for iron, as well as for the description of the magnitude of the Seebeck coefficient, its temperature and composition dependence. 

\end{abstract}
\maketitle

\section{Introduction}
\label{intro}
\indent Cobalt monosilicide and its solid solutions with Fe and Ni were known for a long time as semimetals and were considered as promising thermoelectric materials due to their large Seebeck coefficient and high electrical conductivity \cite{Asanabe1964,Fedorov1995}. 
Recently, interest to these compounds was renewed in connection with the study of their topological properties. The compounds cristallize into B20 crystal structure. The structure  belongs to the space group  \#198  ($P2_13$) that lacks the inversion symmetry. As it was shown recently \cite{Bradlyn2016}, certain space groups allow for the existence of linear band crossing with band degeneracy higher than two. The corresponding nodes can carry non-zero topological charge larger than unity. In ref.~\cite{Bradlyn2016}, these materials where called ``new fermions''. Recent calculations of band structure and topological properties of CoSi \cite{Tang2017,Pshenay2018} and of isostructural compound RhSi \cite{Chang2017} demonstrated that the nodes at the $\Gamma$ and $R$ points of the Brillouin zone of these materials carry topological charges of $\pm 4$, and that the projections of these nodes onto the surface Brillouin zone are connected by Fermi arcs. 
These results indicate that CoSi belongs to the class of topologically non-trivial ``new fermions'' compounds.

The first interpretation of thermoelectric and galvanomagnetic properties of CoSi and its solid solutions with FeSi and NiSi were made using two-band model of electronic spectrum \cite{Asanabe1964,Fedorov1995}. In this model, the band structure near the Fermi level was represented by isotropic parabolic conduction and valence bands that overlap by about 0.02-0.04eV. The density-of-states effective masses of the carriers in these bands were found to be $4-6m_0$ and $2 m_0$ for holes and electrons, respectively. The sign and the magnitude of the Seebeck coefficient in this simple model was determined by the balance of the contributions of electrons and holes. The difference of electron and hole concentrations, determined from the transport coefficients, was of the order of $10^{20}$cm$^{-3}$.

Later, the ab initio calculations of the band structure of these compounds demonstrated that their electronic spectrum is much more complex than this simple model (see, fig.~\ref{fig:atoms_bands}) \cite{Bradlyn2016,Tang2017,Pshenay2018, Pan2007,Sakai2007,Ishii2014}. The first-principle calculations of the Seebeck coefficient of Cr$_{1-x}$Mn$_x$Si, Mn$_{1-x}$Fe$_x$Si, Fe$_{1-x}$Co$_x$Si, and Co$_{1-x}$Ni$_x$Si solid solutions were made in ref.~\cite{Sakai2007} in the constant relaxation time approximation (CRTA). The results reproduce some features of the Seebeck coefficient dependence on  composition and on temperature in these solid solutions, but calculated value of the Seebeck coefficient in CoSi was positive, increasing with temperature, in contrast to experimental data. This discrepancy can be caused by inaccuracies in the description of electronic band structure or scattering. 

In our previous work, we investigated the influence of many body $G_0W_0$ corrections to electronic band structure on the Seebeck coefficient in the constant relaxation time approximation \cite{Pshenay2017}. The results of these calculations were in a reasonable agreement with the experimental data on temperature and composition dependence of the Seebeck coefficient. The calculations showed that, due to small spin-orbital splitting, the account of spin-orbit coupling (SOC) has only a marginal effect on the transport coefficients at not too low temperatures. Finally, it was noted, that there is a possibility for further improvement of the  theoretical description of thermoelectric properties by going beyond CRTA in the description of the electron scattering. 

In the present work, we study the influence of the energy-dependent charge carrier scattering on the transport coefficients of CoSi and Co$_{1-x}$M$_x$Si (M=Fe, Ni) diluted alloys. We consider the most important scattering mechanisms, i.e. the electron scattering by phonons, and by point defects. The scattering rates were obtained from the first principle calculations. 
These calculations confirmed the strong effect of the energy dependence of the relaxation time on the Seebeck coefficient. For some of electronic bands, they revealed large contribution of interband scattering into the total scattering rate. Based on these theoretical results, we analyze the dependence of thermopower and resistivity on temperature and Fe (Ni) content in CoSi-based solid solutions.

\section{Scattering rate calculations}
\label{sec:1}

\begin{figure*}
\includegraphics[width=15cm]{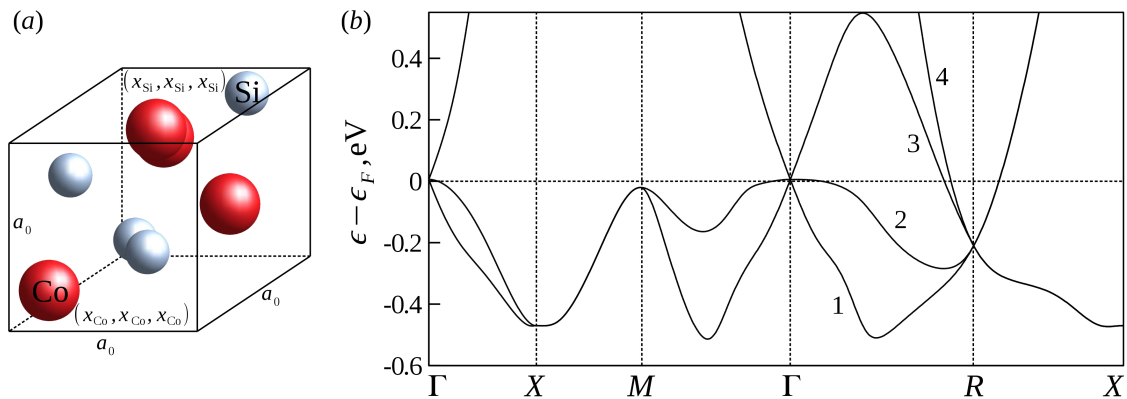}
\centering
\caption{Unit cell of CoSi in cubic B20 crystal structure ($a$). First principle electronic band structure of CoSi without spin-orbit coupling ($b$). Numbers label the bands in the order of their energy increase. Energy $\epsilon$ is measured relative to Fermi level $\epsilon_F$ in CoSi.}
\label{fig:atoms_bands}
\end{figure*}

The ab initio calculations of band structure of CoSi were performed using QuantumEspresso package \cite{QE1,QE2}. They were made using gradient corrected density functional (GGA-PBE) \cite{PBE} without spin-orbit coupling. We used optimized norm-conserving pseudopotentials \cite{ONCV1,ONCV2,ONCV3}, plane wave energy cutoff of 90~Ry and 6x6x6 Monkhorst-Pack (MP) grid in reciprocal space (corresponding total energy convergence was of the order of 1~meV/at). The unit cell of considered material is cubic and contains four formula units (fig.~\ref{fig:atoms_bands}$a$). There are four equivalent atoms of each type (Co or Si) in the unit cell. The positions of these atoms in crystalline coordinates are $(x,x,x)$, $(-x+1/2,-x,x+1/2)$, $(-x,x+1/2,-x+1/2)$ and $(x+1/2,-x+1/2,-x)$, where $x=x_{\mathrm{Co}}$ or $x=x_{\mathrm{Si}}$ (fig.~\ref{fig:atoms_bands}$a$). The optimized lattice constant and parameters, describing Co and Si position in the unit cell, were $a_0=$4.4348$\mathrm{\AA}$ and $x_{\mathrm{Co}}=$0.145, $x_{\mathrm{Si}}=$0.843 that quite good correlate with experimental values $a_0$= 4.4445$\mathrm{\AA}$ and $x_{\mathrm{Co}}=$0.144, $x_{\mathrm{Si}}=$0.846 \cite{Fedorov1995}. The band structure of CoSi in GGA-PBE approximation in the vicinity of the Fermi level is plotted in fig.~\ref{fig:atoms_bands}$b$ and agrees well with the results presented in Refs.~\cite{Pan2007,Sakai2007,Ishii2014,Bradlyn2016,Tang2017,Pshenay2018}.

\begin{figure}
\includegraphics[width=8cm]{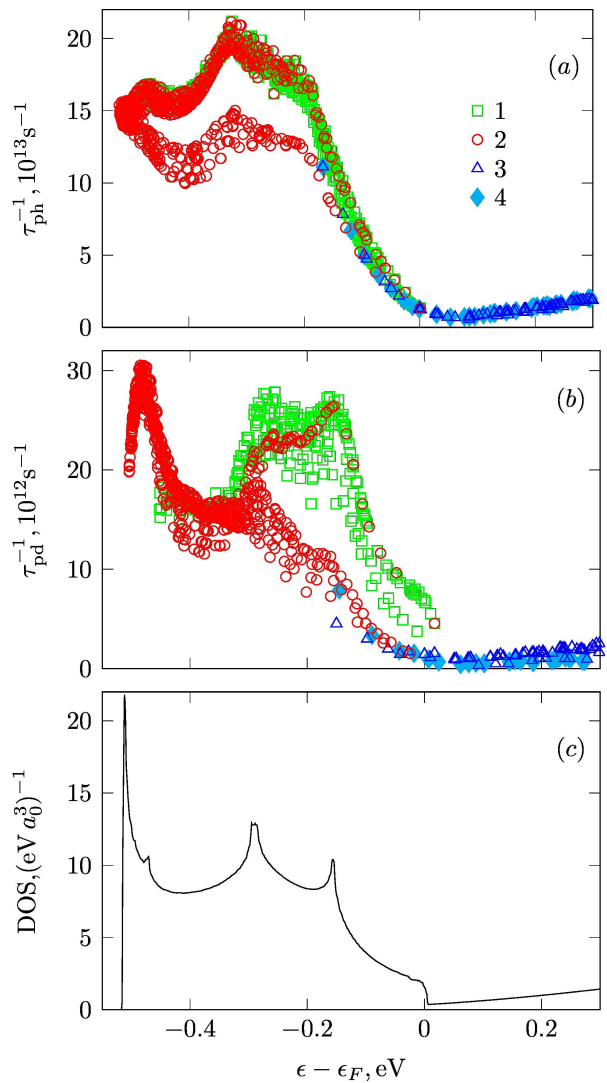}
\centering
\caption{Scattering rate due to electron-phonon ($a$) and short-range impurity scattering ($b$) together with the density of states ($c$). Numbers on legend correspond to the band indexes (see fig.~\ref{fig:atoms_bands}).}
\label{fig:lw-dos}
\end{figure}

The calculation of electron-phonon interaction was performed with the help of EPW software package \cite{Giustino2007,Ponce2016}. At the first step, the phonon dynamic matrices were calculated, using density-functional perturbation theory \cite{Baroni2001}. The calculations were made in the irreducible wedge of the Brillouin zone (BZ) for k-points, corresponding to 6x6x6 $\Gamma$-centered MP-grid. The calculated phonon spectrum well correlates with the spectrum obtained previously, using supercell approach \cite{Pshenay2017}. 
In the course of these calculations, the self-consistent potentials ${\Delta}V_{\mathbf{q} \nu}$, corresponding to phonon perturbations with given wave vectors $\mathbf{q}$ and phonon branches $\nu$, were obtained. That allowed to calculate matrix elements for the electron scattering due to perturbations induced by these phonon modes. EPW uses Wannier representation for both electrons and phonons to interpolate these matrix elements to a finer grid, to calculate the imaginary part of the electronic self-energy $\mathrm{Im}\Sigma$ arising due to the electron-phonon scattering. The imaginary part of electron self-energy is proportional to the scattering rate: $2\mathrm{Im}\Sigma/\hbar = \tau_{\mathrm{ph}}^{-1}$, which corresponds to the energy relaxation time and is used as a first approximation in the calculation of transport properties \cite{Song2017}. The relaxation rate $\tau_{\mathrm{ph}}^{-1}$ for the full k-point mesh is plotted in fig.~\ref{fig:lw-dos}$a$ as a function of energy. The calculations were made on 36x36x36 mesh in BZ for the temperature of 300~K.

Besides phonon scattering, in CoSi and, especially, in solid solutions with Fe or Ni, there should be also point defect scattering. The usually considered scattering mechanisms are the short-range or Coulomb scattering with scattering matrix elements given by:

\begin{equation}
V_{m n}(\mathbf{k}',\mathbf{k}) = \sum_{\mathbf{q}}^{} {V(\mathbf{q}) \delta_{
 \mathbf{k}+\mathbf{q}, \mathbf{k}'+\mathbf{G}
}}
\langle u_{m\mathbf{k}'} | u_{n\mathbf{k}} \rangle
\label{eq:Mmn}
\end{equation}

\noindent where $V(\mathbf{q})$ is the Fourier transform of scattering potential, $u_{n\mathbf{k}}$ are Bloch amplitudes of electronic state with wave vector $\mathbf{k}$ in band $n$ and $\mathbf{G}$ is reciprocal lattice vector \cite{Ridley2013Quantum}. 

For single defect at the point $\mathbf{r}_0$ in the case of short-range scattering potential $V(\mathbf{r}-\mathbf{r}_0) = U_0 \delta(\mathbf{r}-\mathbf{r}_0)$, the square modulus of the Fourier component is independent of $\mathbf{q}$ ($|V(\mathbf{q})|^2=U_0^2$). For the case of dilute alloys, considered here, we use independent scattering approximation, for which scattering probability does not depend on $\mathbf{r}_0$ and is proportional to the defect concentration \cite{Ridley2013Quantum}. The strength of the short-range potential was estimated using supercell approach for cubic 2x2x2 supercell consisted of eight cubic unit cells (64 atoms). We calculated self-consistent potential $V_0$ of the CoSi supercell and self-consistent potential $V_d$ for the same supercell with one of Co atoms replaced by Fe or Ni. The supercell with such substitutional defect corresponds to the solid solution Co$_{1-x}$M$_x$Si (M = Fe or Ni) with $x\approx0.03$. The strength of the scattering potential, obtained as the real space integral of the difference $V_d-V_0$, was equal to $|U_0|=$14.2 and 6.4~eV\AA$^3$ for Fe and Ni, respectively. Due to periodic boundary conditions for supercell and the equivalence of all Co positions, the result is independent of which Co atom was replaced by Fe (Ni). The values of $|U_0|$ obtained allowing for atomic relaxation within the supercell differ only by about 1\% from the values given above.  

For Coulomb scattering, $V(q)=4 \pi e^2 r_{\mathrm{scr}}^2 /((q\,r_{\mathrm{scr}})^2+1)$ \cite{Ridley2013Quantum}. Since  CoSi and its alloys are almost metallic materials, it is assumed here that the static dielectric constant is equal to unity \cite{Mott1936}. The calculated density of states near the Fermi level is about $g(\epsilon_F)\approx 0.023 \mathrm{eV}^{-1}\,\mathrm{\AA}^{-3}$. Thus, the Thomas-Fermi screening radius is about $r_{\mathrm{scr}} = (4 \pi e^2 g(\epsilon_F))^{-1/2} \approx 0.5$~\AA. Due to such strong screening, the scattering probability varies slowly with $q$ and is not small even for large $q$, comparable with the dimension of BZ. The estimation gives the Coulomb scattering potential equal to $4 \pi e^2 r_{\mathrm{scr}}^2\approx$43.6~eV\AA$^3$ for small $q$. 

The overlaps of Bloch amplitudes of initial and final states (see eq.(\ref{eq:Mmn})) were calculated with the help of QuantumEspresso for 25x25x25 BZ-mesh. The corresponding rate of elastic scattering on short-range potential is plotted in fig.~\ref{fig:lw-dos}$b$ as a function of energy. The calculations were made, assuming independent impurity scattering, for $x\approx0.03$ and with $U_0$ parameter for Fe substitution. Due to small $r_{\mathrm{scr}}$, the energy dependence of the scattering rate for Coulomb scattering (not shown here) is very similar to the dependence, presented in fig.~\ref{fig:lw-dos}$b$ for short-range scattering.  However they differs in magnitude, according to the different strengths of corresponding potentials. Therefore, in the following calculations, we  assume that the energy dependence of relaxation time for point defect scattering is the same as corresponding dependence for short-range impurity scattering. The strength of the scattering potential will be obtained from the low-temperature resistivity values.

The scattering rate depends, among other factors, on the density of final states \cite{Ridley2013Quantum}. In the simple case of a single isotropic parabolic band and nearly elastic acoustic scattering, or short range impurity scattering, the inverse relaxation time is proportional to the density of states in considered band \cite{Ridley2013Quantum}. If electron transitions to another energy band are allowed, the scattering rate will be proportional to the total density of the final states. In general case, the scattering rate will depend also on matrix elements of electronic transitions and, in the case of electron-phonon scattering, on phonon energies and distribution function. The scattering rates in fig.~\ref{fig:lw-dos} include all these factors. It can be seen that the dependence of scattering rate on energy is very different for energies above and below $\epsilon_F$. 
This energy dependence correlates with the energy dependence of the total DOS (fig.~\ref{fig:lw-dos}$c$), implying the importance of the interband scattering for bands with the low initial density of states.  

\begin{figure}
\centering
\includegraphics[height=4.5cm]{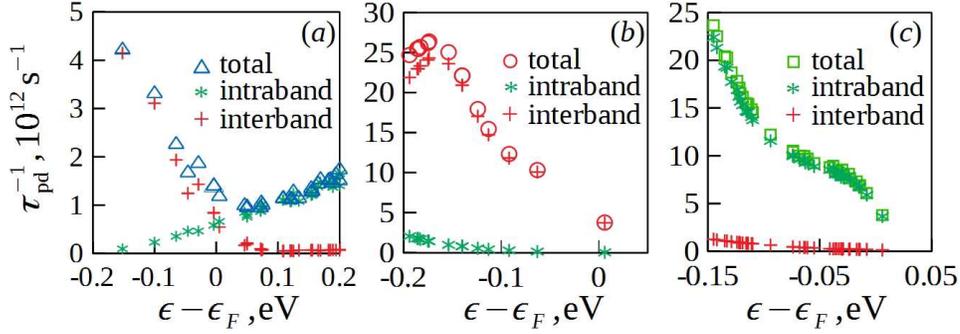}
\caption{Contributions of intraband and interband scattering to the total point defect scattering rate for electronic states in the band \#3 near $R$ point ($a$) and in the bands \#1 ($b$) and \#2 ($c$) near $\Gamma$ point.}
\label{fig:wshr}
\end{figure}

For the case of point defect scattering, it is easy to separate the interband and  intraband contributions. Fig.~\ref{fig:wshr} illustrates relative contributions of the intraband and the interband scattering for different bands in a vicinity of the $R$ (fig.~\ref{fig:wshr}$a$) and $\Gamma$ (fig.~\ref{fig:wshr}$b$-$c$) points. In these figures the intraband scattering corresponds to electron transitions with both the initial and final states belonging to the same band, and with both the initial and final wave vectors lying in a vicinity of the same point in BZ (i.e., restricted by inequality $|k_i-k_{P i}|<\pi/2 a_0$, where $i=x,y,z$ and $P=\Gamma$ or $R$). It can be seen that for $\epsilon>\epsilon_F$ for states in the 3rd band near $R$ point (fig.~\ref{fig:wshr}$a$) the main contribution to the scattering comes from the intraband transitions. The intraband scattering rate decreases with decreasing energy due to the reduction of the density of states in this band near $R$ point. For $\epsilon<\epsilon_F$, $\tau_{pd}^{-1}$ increases with decreasing energy.  The main contribution to the total scattering rate in this energy range  comes from the interband transitions. The energy dependence of the scattering rate correlates with the energy dependence of the total density of states. Similar behavior is found for 4-th band near $R$ point, the results are shown in fig.~\ref{fig:lw-dos}$b$.

\begin{figure}
\centering
\includegraphics[height=4.5cm]{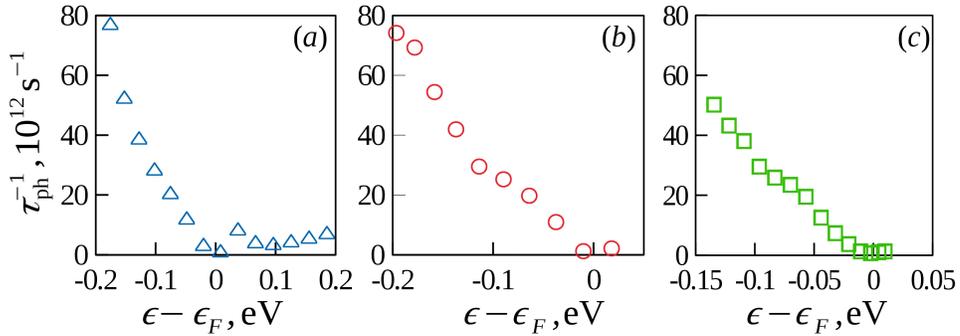}
\caption{Total scattering rate due to electron-phonon interaction for electronic states in the band \#3 near $R$ point ($a$) and in the bands \#1 ($b$) and \#2 ($c$) near $\Gamma$ point.}
\label{fig:wepw}
\end{figure}

For the first band near $\Gamma$ point, the interband scattering is important for the whole considered energy range (fig.~\ref{fig:wshr}$b$). It is dominated by the transitions to the second band with weak dispersion and, hence, with  large DOS. On the other hand, for the second band near $\Gamma$ point, the intraband scattering prevails (fig.~\ref{fig:wshr}$c$), because 2-nd band gives the main contribution to the total DOS. This is illustrated in fig.~\ref{fig:lw-dos}$b$, where just below Fermi energy, three distinct curves for $\tau_{\mathrm{pd}}^{-1}$ can be separated. The similarity to the energy dependence of the total DOS is particularly evident for the upper one, which corresponds to the scattering rates for the states in two lowest bands near $\Gamma$ point. The scattering rates are mainly determined by the interband scattering from the 1st to the 2-nd band and by the intraband scattering inside the 2-nd band near $\Gamma$ point. Two lower series of points correspond to scattering rates for the states in bands \#2 and \#1, 3, 4 respectively, located far from the $\Gamma$ point. 

For the electron-phonon scattering, the matrix elements of electron transitions in the Bloch representation are not easily available from the EPW code, so the detailed analysis of the interband and the intraband transitions was not performed here. But the scattering rates for a specific band and special points in the BZ can be extracted. 
They are shown in fig.~\ref{fig:wepw}. 
Comparing figs.~\ref{fig:wshr} and \ref{fig:wepw}, one can see the similarity in the energy dependence of the total scattering rate for both scattering mechanisms. Thus, we can suggest that for electron-phonon scattering the separation of total scattering rate into intraband and interband contributions should be very similar to the case of point defect scattering, discussed above. 
For example, the contribution of interband transitions to the total scattering intensity is quite important for band \#3 near $R$ point for $\epsilon<\epsilon_F$ (fig.~\ref{fig:wepw}$a$). Otherwise, for electrons close to $R$ point, the transition probability would decrease when the energy decreases below $\epsilon_F$. Also the scattering rates for the 1st and 2nd bands near $\Gamma$ point are very similar (fig.~\ref{fig:wepw}$b,c$). It can be explained assuming that both scattering rates are determined by the large density of states in the 2nd band, thus, implying the importance of interband scattering for the 1st band and intraband scattering for the 2nd band. 

Comparing plots in fig.~\ref{fig:lw-dos}$a$-$c$, we can conclude that for both considered scattering mechanisms, the scattering rate is stronger below $\epsilon_F$ than above it, and that its energy dependence correlates with the energy dependence of total DOS. 
As will be shown below, it will strongly influence the magnitude of the Seebeck coefficient and the dependence of conductivity and thermopower on the doping level.

\section{Calculation of transport coefficients}
\label{sec:2}

In order to calculate the electrical conductivity and Seebeck coefficient we first obtain transport distribution function (TDF):

\begin{equation}
\Xi_{i j}(\epsilon) = \frac{2}{V} \sum_{n \mathbf{k}} {
    v_i(n, \mathbf{k}) v_j(n, \mathbf{k}) 
    \tau(n, \mathbf{k}) 
    \delta(\epsilon-\epsilon(n, \mathbf{k})) 
},
\label{eq:TDF}
\end{equation}

\noindent where $\epsilon(n, \mathbf{k})$, $v_i(n, \mathbf{k})$ and $\tau(n, \mathbf{k})$ are energy, velocity projection on Cartesian axis $i$ and relaxation time for electron in the $n$-th band at the point $k$ in the Brillouin zone, respectively. 
The $\delta$-function integration was performed on 96x96x96 MP-grid using optimized tetrahedra integration \cite{OptTetra}. The relaxation times obtained above were interpolated onto this fine grid using  smoothed Fourier interpolation as implemented in BoltzTrap2 \cite{BoltzTrap2}. In cubic materials, TDF becomes diagonal tensor, so in the following we drop Cartesian indexes. Then the electrical conductivity $\sigma$ and the Seebeck coefficient $S$ can be calculated as

\begin{equation}
\sigma = e^2 \int_{-\infty}^{\infty}{
  d\epsilon (-f_0'(\epsilon, \mu, T)) \Xi_{i i}(\epsilon)
},
\label{eq:sigma}
\end{equation}

\begin{equation}
S = \frac{e}{\sigma T} \int_{-\infty}^{\infty}{
  d\epsilon (-f_0'(\epsilon, \mu, T)) (\epsilon-\mu) \Xi_{i i}(\epsilon)
},
\label{eq:S}
\end{equation}
\noindent where $\mu$ is the chemical potential and $f_0'(\epsilon, \mu, T)$ is the energy derivative of Fermi-Dirac distribution function.

In the present work we use the rigid band approximation in order to account for the chemical potential shift due to the substitution of Co by Fe(Ni) or due to the temperature change. In the course of our calculations, we use $\epsilon_F$, the Fermi level in CoSi at 0~K, as the origin of the energy. The chemical potential $\mu$ is also counted from the same origin. In Co$_{1-x}$Ni$_x$Si, the atomic fraction of nikel $x$ is equal to the relative change of electronic concentration $\delta n = (n-n_0)/n_0$, if we define $n_0$ and $n$ to be the electron concentration in four considered electronic bands of CoSi and solid solutions, respectively. In this case, the chemical potential $\mu$ of the solid solution is greater than the Fermi energy $\epsilon_F$ of CoSi. For Fe substitution $\delta n = -x$ and $\mu<\epsilon_F$.

\begin{figure}
\centering
\includegraphics[width=7cm]{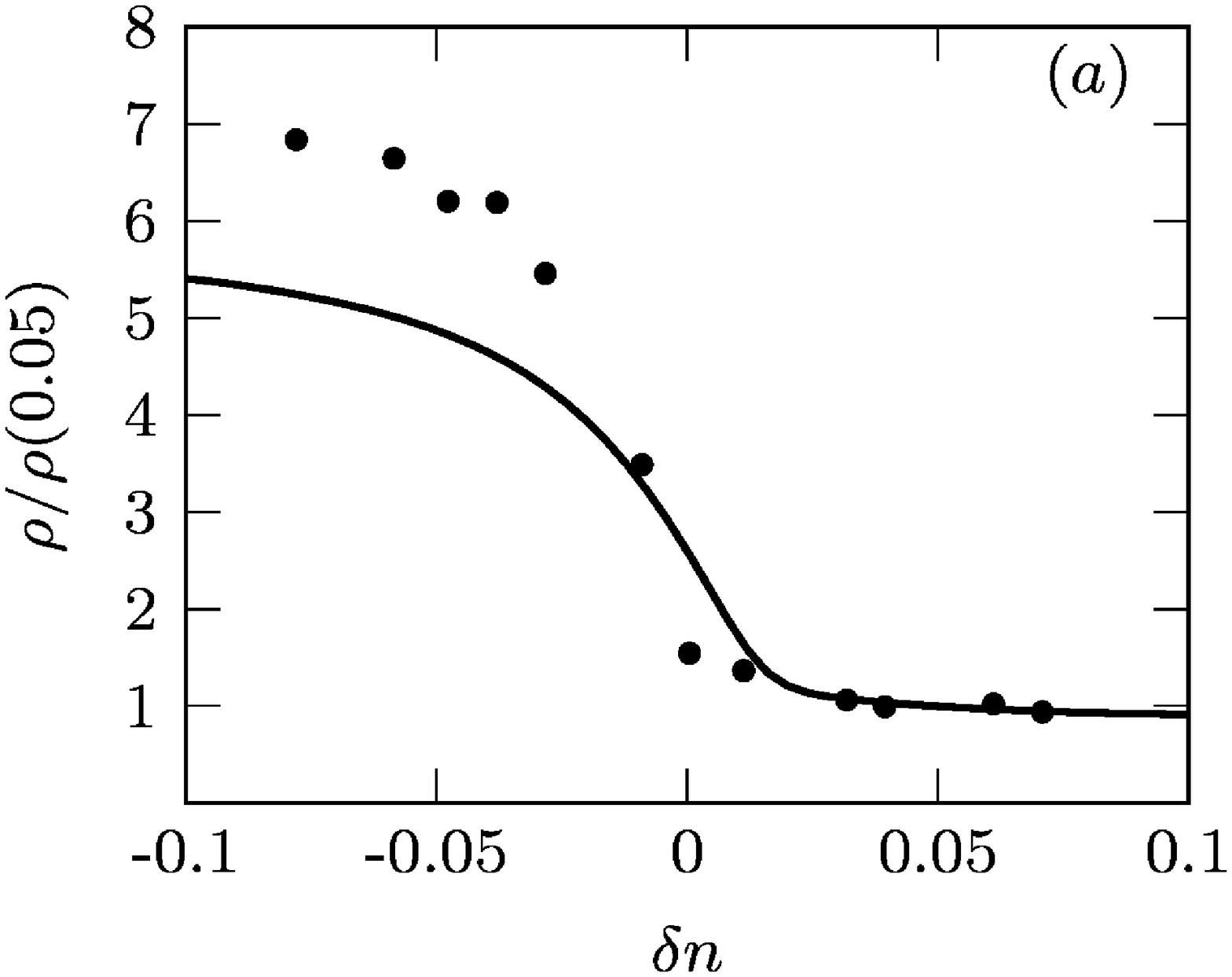} 
\includegraphics[width=7.5cm]{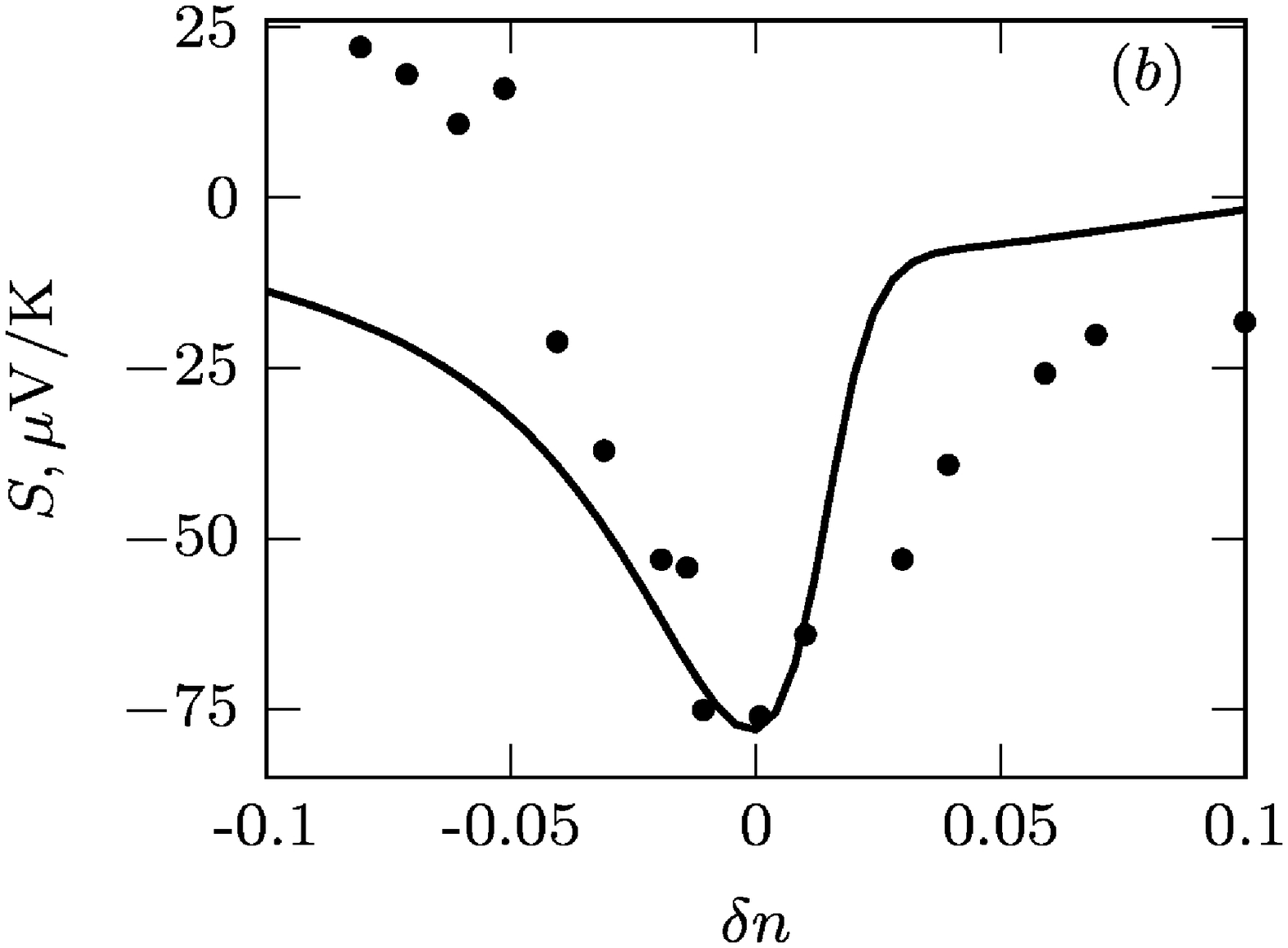}
\caption{The normalized resistivity ($a$) and Seebeck coefficient ($b$) at room temperature on relative change of electron concentration (alloy composition). Symbols correspond to the experimental data \cite{Alekseeva1981_tr}, the lines are the calculation results.}
\label{fig:vs-x}
\end{figure}

At first we calculate the electrical conductivity and Seebeck coefficient as a function of the Co$_{1-x}$Ni(Fe)$_x$Si alloy composition at room temperature. 
One can reasonably expect that at room temperature, the phonon scattering is the most effective scattering mechanism in pure CoSi and in the dilute alloys.
Therefore, as a first step in these calculation we take into account only the electron-phonon scattering.
The alloying effect is restricted only to the chemical potential shift, as it was explained above.
The theoretical resistivity ($\rho = 1/\sigma$) and the Seebeck coefficient are plotted in fig.~\ref{fig:vs-x} in dependence on composition  together with the experimental data from ref.~\cite{Alekseeva1981_tr}. The curves in fig.~\ref{fig:vs-x} were obtained without any fitting parameters. It can be seen that calculations very well reproduce the value of the Seebeck coefficient for CoSi and the experimental fact that the maximum of the absolute value of $S$ is reached at the stoichiometric CoSi composition. 

\begin{figure}
\centering
\includegraphics[width=8.25cm]{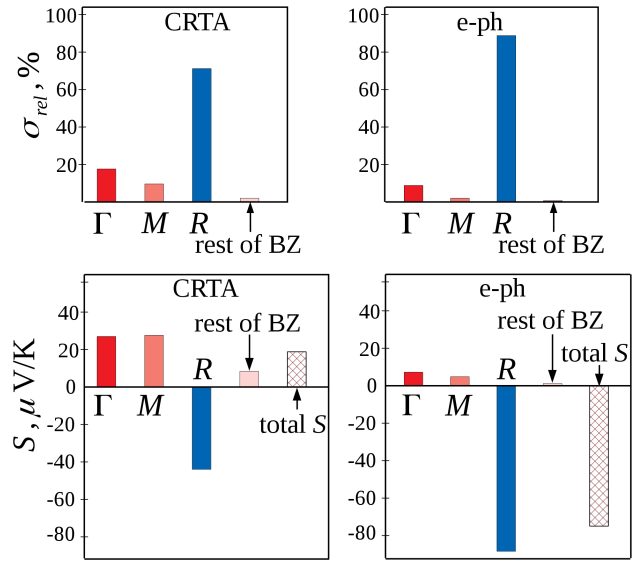}
\caption{Contributions of electrons from states in the vicinity special points in the Brillouin zone ($\Gamma$, $R$ and $M$) to the total electrical conductivity $\sigma_{rel}$ (relative contributions) and the Seebeck coefficient $S$ in CoSi. Left and right columns correspond to calculations performed in CRTA approximation and taking into account electron-phonon scattering at 300K.}
\label{fig:chart}
\end{figure}

Using restrictions on the band index or wave vector values in eq.~(\ref{eq:TDF}), it is possible to separate contribution to the transport coefficients from specific band in certain region of the Brillouin zone. For example, restricting $|k_i-k_{P i}|<\pi/2 a_0$ $(i=x,y,z)$ for special points $P=\Gamma,R, M$, we separated contributions to electrical conductivity and the Seebeck coefficient from states in vicinity of these points in BZ at room temperature (fig.~\ref{fig:chart}). It can be seen that the states close to these points contribute more than 98\% to the conduction. The sates near $R$ point give negative contribution to the Seebeck coefficient, while the states near $\Gamma$ and $M$ points in the BZ give positive contributions. 

In constant relaxation time approximation, the balance of these contributions depends only on the chemical potential level that can be changed by the alloy composition.  
The DFT band structure predicts that in CoSi the chemical potential lies just below the node at the $\Gamma$ point in hole-like band. The concentration of electrons in bands \#3-4 near $R$ point is $4.13\cdot10^{20}$cm$^{-3}$ and hole concentration in bands \#1-2 near $\Gamma$ and $M$ points is a little bit larger but similar $4.2\cdot10^{20}$cm$^{-3}$. Due to larger DOS in hole-like bands, holes in these bands are less degenerate than electrons in bands \#3-4 close to the $R$ point. The absolute value of the Seebeck coefficient decreases with the increase of degeneration. Thus, the Seebeck coefficients for states near $\Gamma$ and $M$ points are 155 and 291$\mu$V/K, that is larger in absolute values than the contribution from states near $R$ point (-62$\mu$V/K). Total Seebeck coefficient can be obtained as a sum of these parts, weighted by relative contributions to total conductivity (in fig.~\ref{fig:chart}, such weighted contributions to thermopower are shown). Though the contribution to conductivity from $\Gamma$ and $M$ points are rather small (fig.~\ref{fig:chart}), due to large Seebeck coefficient the contribution of holes prevails and results in small positive $S$ values ($\approx$19$\mu$V/K) in contradiction with the experiment. The calculations \cite{Sakai2007} showed, that at higher temperatures this positive Seebeck coefficient increases in magnitude. 

Manybody G$_0$W$_0$ corrections to the band structure \cite{Pshenay2017} shift relative energy positions of bands and chemical potential and partially improve results in the CRTA approximation. 
But the minimum in the concentration dependence of thermopower was predicted at small positive $\delta n$ instead of $\delta n=0$ and was smaller in absolute value compared to the experiment. 

The results of the present calculations suggest that more important reason for this dependence of $S$ on composition is the specific energy dependence of the relaxation time, inversely proportional to the total density of states.
In general, if carriers with higher energies scatter stronger than the ones with lower energies, the Seebeck coefficient becomes smaller in comparison with the case of energy independent scattering (CRTA) and vice versa. 
It is especially prominent for degenerate electron systems because, in this case, charge carriers with $\epsilon>\mu$ and $\epsilon<\mu$ give contributions of different signs to the Seebeck coefficient. For example, in the electron-like bands near $R$ point, electrons below the chemical potential give positive contribution to the Seebeck coefficient and the ones above $\mu$ give negative contribution (see eq.~(\ref{eq:S})). 
Prevailing scattering of electrons with $\epsilon<\mu$ reduces the positive contribution to $S$, making the Seebeck coefficient more negative (-99 instead of -62$\mu$V/K in CRTA).
For holes in the vicinity of $\Gamma$ and $M$ points, the scattering rate in our calculations increases with 
increasing of hole energy and the partial Seebeck coefficients becomes smaller, in comparison with the CRTA approximation. It decreases from 155 to 83$\mu$V/K (from 291 to 242$\mu$V/K) for the holes near $\Gamma$ ($M$) point. 
In addition to this effect the relative contribution to the total conductivity from the states near $R$ point increases, while the relative contributions of the stats near $\Gamma$ and $M$ points decreases (see fig.~\ref{fig:chart}).
As a result the total thermopower becomes negative and large in the absolute value ($\approx$-75$\mu$V/K). Interestingly, this is achieved at the specific position of chemical potential, such that the  scattering rates below and above it differ strongly. 
In Co$_{1-x}$M$_x$Si alloys $\mu$ moves to the energy region were DOS depends weaker on energy and the Seebeck coefficient decreases in absolute value for both Ni and Fe substitution, fig.~\ref{fig:vs-x}.

It can be noted that the influence of interband scattering on the thermoelectric efficiency $ZT$ was considered previously on the base of a phenomenological model, both for semiconductors and semimetals \cite{Pshenay2010}. It was shown that in semiconductors with two conduction bands, the interband scattering reduces thermoelectric efficiency. Whereas in semimetals the possibility of interband scattering can lead to the increase of $ZT$. Here, using ab initio calculations, we showed that in CoSi this mechanism is responsible for the high thermopower of this semimetal.

While the obtained relaxation time due to electron-phonon scattering quite well reproduces the behavior of the Seebeck coefficient, it gives too high values for electrical conductivity, compared to experiment. It can be partially connected with the accuracy of the method itself and in part is connected with the presence of point defects. For CoSi at room temperature, experimental value of electrical conductivity is about 7500 S/cm, while calculations with the account of only phonon scattering results in 10660~S/cm, that is about 1.4 times larger. This deviation is typical for this method of calculations \cite{Giustino2007,Ponce2016,Song2017}. Therefore fig.~\ref{fig:vs-x}$a$ shows the
concentration dependence of resistivity normalized to resistivity of Co$_{0.95}$Ni$_{0.05}$Si solid solution.
It can be seen that even the relative change of resistivity can almost be reproduced with this approach. Thus, the asymmetry of composition dependence of resistivity is connected with the asymmetry of the density of states and scattering intensity. The deviation from experiment in fig.~\ref{fig:vs-x}$a$ is connected with additional influence of point defect scattering, which is stronger in alloys with Fe substitution. This behavior cannot be even qualitatively reproduced in CRTA approximation. CRTA approximation predicts the decrease of resistivity because of the increase of electron or hole concentration due to Ni or Fe doping.

\begin{figure}
\centering
\includegraphics[width=7cm]{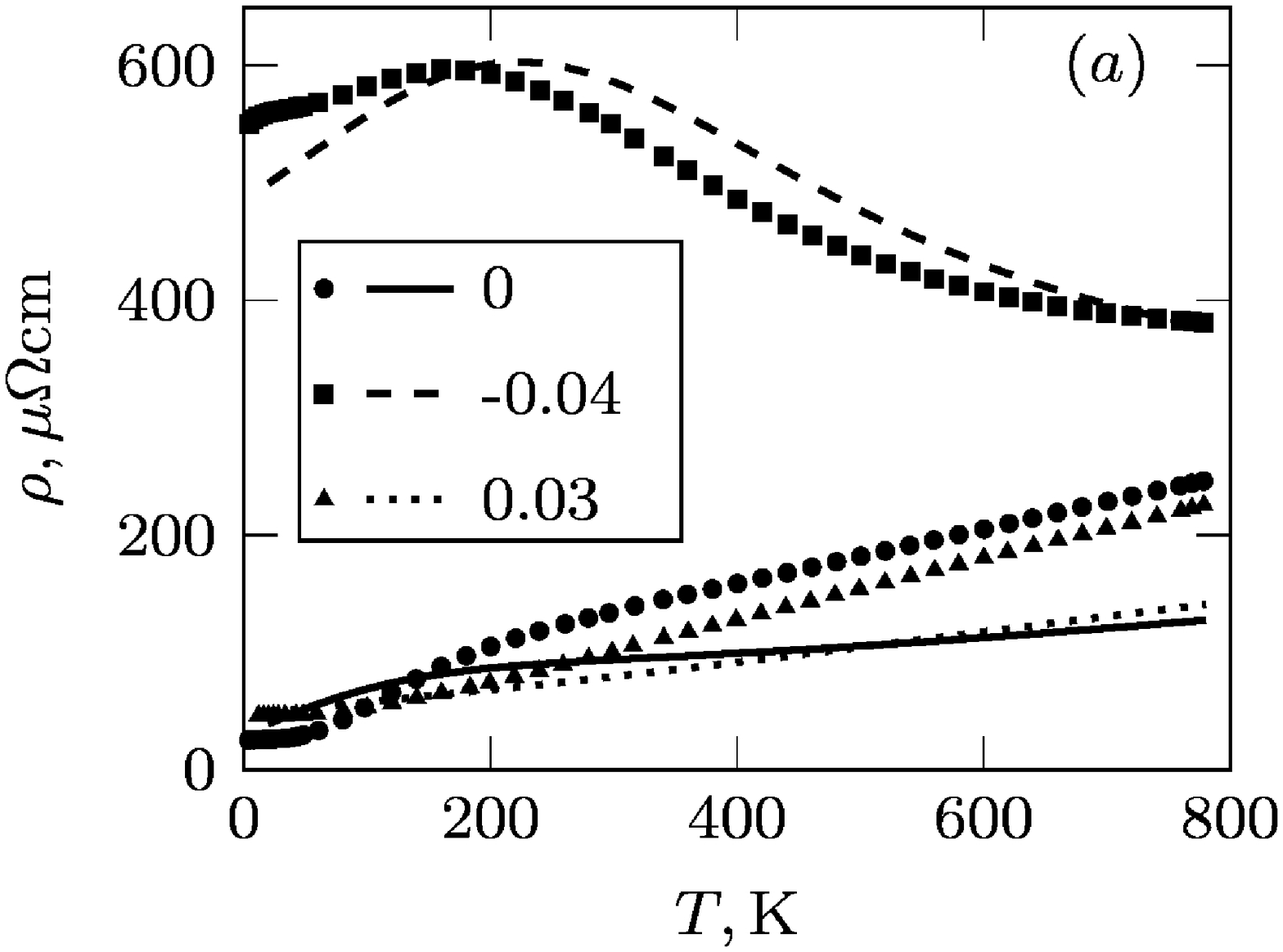} 
\includegraphics[width=7.4cm]{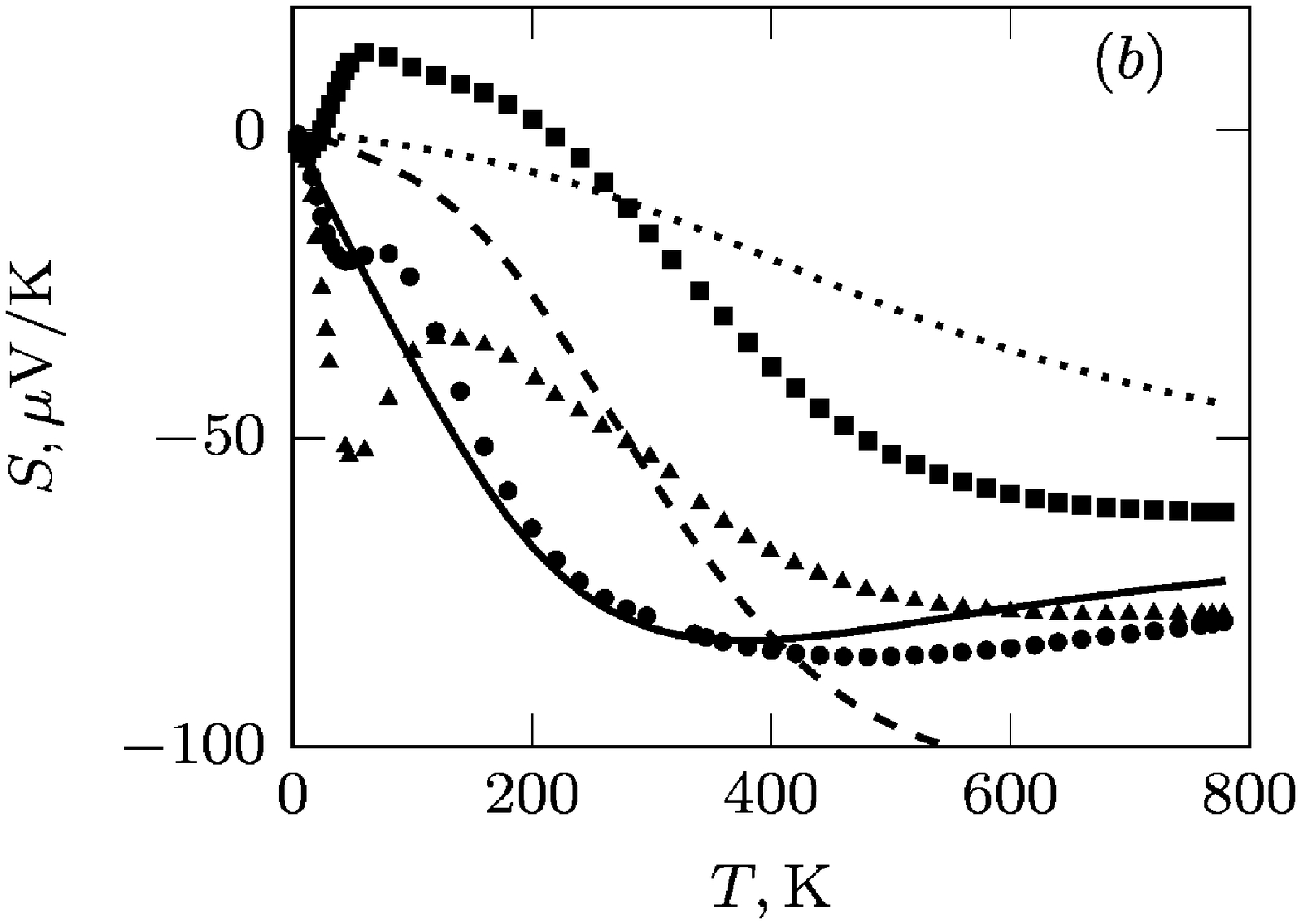}
\caption{Temperature dependences of resistivity ($a$) and thermopower ($b$) for samples CoSi ($\delta n=0$), Co$_{0.96}$Fe$_{0.04}$Si ($\delta n=-0.04$) and Co$_{0.97}$Ni$_{0.03}$Si ($\delta n=0.03$). 
Each line in the legend corresponds to one of these compositions. The circles, squares and triangles are the experimental data from \cite{BurkovECT2017}. The lines are the results of present calculations.}
\label{fig:vs-T}
\end{figure}

\begin{figure}
\includegraphics[width=8cm]{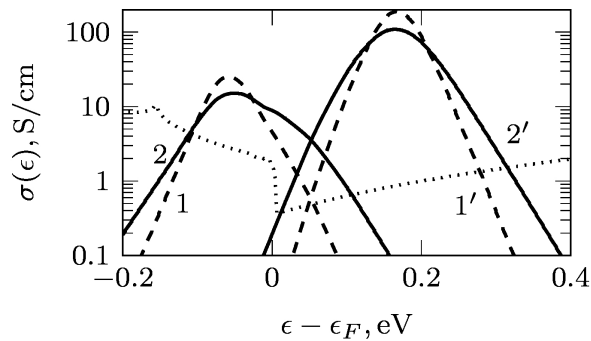}
\centering
\caption{Contribution to electrical conductivity from states with different energy for Co$_{0.96}$Fe$_{0.04}$Si (1, 2) and Co$_{0.97}$Ni$_{0.03}$Si (1', 2'). Dashed lines (1, 1') correspond to $T=$200~K and solid lines (2, 2') - to $T=$300~K. Dotted line shows density of states in arbitrary units. Energy $\epsilon$ is measured relative to $\epsilon_F$ in pure CoSi.}
\label{fig:sigma-ene}
\end{figure}

At temperatures below 50K, the point defect scattering prevail. 
The value of the scattering potential of point defects $|U_0|=$70~eV\AA$^3$ (59~eV\AA$^3$) for Fe (Ni) substitution (at $x$=0.03) was obtained from experimental data on low-temperature resistivity at 50K \cite{BurkovECT2017}. These values seems to be reasonable, assuming scattering on combination of screened Coulomb and short range potential. At intermediate temperatures, the relaxation time was calculated as $\tau^{-1} = \tau_{pd}^{-1} + \tau_{ph}^{-1}(300)\cdot (T/300)$. This simplified relationship was chosen in order to accelerate computer calculations, but it allowed to qualitatively address some peculiarities of the temperature dependence of resistivity (fig.~\ref{fig:vs-T}). From fig.~\ref{fig:vs-T}$a$, it can be seen that, in CoSi and solid solutions with Ni, resistivity above 200~K increases approximately linearly with temperature. It is expected behavior because the electron-phonon scattering rate increases proportional to the temperature at not too small $T$. In the solid solutions with Fe, the resistivity at low temperatures is about an order of magnitude greater and decreases with temperature above 200~K. This unexpected behavior is connected with the fact that point defect scattering in the presence of Fe is stronger not only due to larger scattering potential strength, but mainly due to the fact that in this case the chemical potential is situated at the region of the high DOS and the scattering rates. In this case, the increase of resistivity with the temperature due to phonon scattering is compensated by the redistribution of charge carriers over wider energy range. Due to this redistribution, at higher temperatures more charge carriers are excited  to the high energy region of spectrum, where DOS is smaller and carrier mobility is higher. This is illustrated in fig.~\ref{fig:sigma-ene}, where the contribution of the charge carriers with different energy to the conductivity is plotted. The figure shows the integrand of eq.(\ref{eq:sigma}). These carrier distributions are compared at 200~K (dashed curves) and at 300~K (solid curves). In the case of Ni doping (curves 1', 2'), the increase of temperature leads to the decrease of the maximum of this distribution and its symmetric widening. But in the case of Fe doping (curves 1, 2), the distribution is expanding mainly to the high energy region, where DOS is smaller. Thus, relative contribution of high mobility carriers increases and the total resistivity decreases.

The temperature dependence of the Seebeck coefficient is shown in fig.~\ref{fig:vs-T}$b$. It can be seen that, for CoSi, the calculations quite well reproduce experimental data. For alloys, the general behavior is similar, but the magnitude of thermopower deviate strongly, suggesting that further improvement of the model is necessary. The negative peaks of thermopower at about 50K cannot be explained in the present model as they probably connected with the phonon drag contribution \cite{Pshenay2017}.

\section{Conclusions}
\label{sec:3}

In the present work, we consider the effect of the electron-phonon and point defect scattering on the thermoelectric transport in CoSi and in dilute Co$_{1-x}$M$_x$Si (M= Fe or Ni, $x<0.1$) solid solutions in rigid band approximation. The electron-phonon scattering rate in CoSi was calculated from first principles. The energy dependence of the point defect scattering rate was calculated using ab initio electronic spectrum and wave functions. The strength of the scattering potential was obtained from the low-temperature resistivity.
It was shown that, for both scattering mechanisms, the energy dependence of the scattering rates correlate with that for the total density of states, implying that the contribution of the iterband scattering for electrons from certain bands in this material is important. Strong energy dependence of the relaxation time leads to enhancement of the Seebeck coefficient in comparison with the constant relaxation time approximation. 
Calculated temperature dependence of the Seebeck coefficient in CoSi is in a good agreement with experimental results. The theory gives correct description of evolution of the Seebeck coefficient and of the relative change of the electrical resistivity with composition of dilute Co$_{1-x}$M$_x$Si solid solutions. It was shown that point defect scattering and phonon scattering of charge carriers is stronger in alloys with Fe than with Ni due to stronger scattering potential and larger density of states at the Fermi level. The difference of the temperature dependences of the resistivity in Ni and Fe doped materials was explained. The observed decrease of the resistivity with temperature in the Fe-doped CoSi is connected with the increase in the contribution of carriers with higher energy and larger mobility.

\section*{Acknowledgments}
The study was supported by the Russian Science Foundation, Project No. 16-42-01067.

\section*{References} 
\bibliography{bibl}

\end{document}